\documentclass[%
 reprint,
 amsmath,amssymb,
 aps,
prl,
]{revtex4-1}

\usepackage{graphicx}
\usepackage{dcolumn}
\usepackage{bm}
\usepackage{graphicx}
\usepackage{hyperref}
\usepackage{natbib}

\begin{document}

\preprint{APS/123-QED}

\title{The Mechanism Behind Beauty: Golden Ratio Appears in Red Blood Cell Shape}
\author{Xue-Jun Zhang}
\author{Zhong-Can Ou-Yang}%
 \email{oy@itp.ac.cn}
\affiliation{CAS Key Laboratory of Theoretical Physics, Institute of Theoretical Physics, Chinese Academy of Sciences, P. O. Box 2735, Beijing 100190, China.\\
School of Physical Sciences, University of Chinese Academy of Sciences, No.19A Yuquan Road, Beijing 100049, China.}
\date{\today}

\begin{abstract}
  In the past two decades, under the conditions that both the osmotic pressure $\Delta p$ and tensile stress $\lambda$ equal zero, a rigorous solution (RS) of human red blood cell (RBC) with a minus spontaneous curvature $c_{0}$ has been derived with Helfrich model. And, by fitting with observed shapes of RBC, $c_{0}R_{0}$ has been predicted to be -1.62 as minus golden ratio, where $R_{0}$ is the radius of a sphere with the same area of RBC. In this Lett., it is also found $\rho_{max}$ /$\rho_{B}\approx$ 1.6 shows a approximately beautiful golden cross section of RBC, where $\rho_{max}$ is the radius of RBC and $\rho_{B}$ is the radius at maximal thickness of RBC. With a complete numerical calculation, we find the mechanism behind the beauty that minus golden ratio of $c_{0}R_{0}$ is the balance between economical surface area and enough  deformability to pass spleen, the so called ``physical fitness test".

\end{abstract}

\maketitle


Scientists never stop their exploration in red blood cell (RBC) since Jan Swammerdam in 1658 and Anton van Leeuwenhoek observed red balls by microscope a few years later \cite{hamly1947fluorescence,evans1952red,guest1963red,skalak1969deformation}. In 1992, Ou-Yang \textit{et al.} \cite{zhong1992determination}found that multiply RBC's spontaneous curvature by the radius of a sphere with the same area of RBC is negative golden ratio by using their theory and experimental data of Evan Evans \cite{evans1972improved}. The study of golden ratio begins from mathematicians but is not confined just to them. Biologists, artists, musicians, historians, architects, psychologists, and even mystics have pondered and debated the basis of its ubiquity and appeal. In fact, it is probably fair to say that the golden ratio has inspired thinkers of all disciplines like no other number in the history of mathematics \cite{livio2008golden}. In the past 24 years, however, nobody has explained the reason why it happens in the RBC. Our work is trying to give a reasonable explanation for it and predicts for other mammalian erythrocytes.

In 1973, Helfrich established the basic model and theory by introducing spontaneous curvature to biological vesicle with following shape energy \cite{helfrich1973elastic}
\begin{equation}\label{EQ1}
F_{b}=\frac{k}{2}\ointop(c_{1}+c_{2}-c_{0})^{2}dA+\bar{K}\oint c_{1}c_{2}dA,
\end{equation}
\noindent where $k$, $\bar{K}$, $c_{1}$, $c_{2}$, and $c_{0}$ are the bending rigidity, Gaussian-curvature modulus, two principal curvatures, and the spontaneous curvature, respectively. Later, in 1987 Ou-Yang Zhong-Can and Helfrich derived from (\ref{EQ1}) a more general shape equation \cite{zhong1989bending}
\begin{equation}\label{EQ2}
\varDelta p-2\lambda H+k(2H+c_{0})(2H^{2}-2K-c_{0}H)+2k\nabla^{2}H=0,
\end{equation}
\noindent where H and K are the mean and the Gaussian curvatures respectively, and $\nabla^{2}$ is the Laplace-Beltrami operator. $\Delta p$ and $\lambda$ are regarded as the the Lagrange multipliers which take account of the constraints of constant volume and area of the vesicles respectively, $\Delta p$ = $p_{o}$-$p_{i}$, is also the osmotic pressure difference between outer and inner media, and $\lambda$ is the tensile stress.
In 1993, Hu.J.G \textit{et al.} derived the shape equation for an axiymmetrical vesicle from (\ref{EQ2}) \cite{jian1993shape}
\begin{eqnarray}\label{EQ3}
&&{} \notag \cos^{3}\varphi\left(\frac{d^{3\varphi}}{dp^{3}}\right)=4\sin\varphi\cos^{2}\varphi\left(\frac{d^{2}\varphi}{d\rho^{2}}\right)\left(\frac{d\varphi}{d\rho}\right)\\ \notag
&&{}-\cos\varphi\left(\sin^{2}\varphi-\frac{1}{2}\cos^{2}\varphi\right)\left(\frac{d\varphi}{d\rho}\right)^{3}\\ \notag
&&{}+\frac{7\sin\varphi\cos^{2}\varphi}{2\rho}\left(\frac{d\varphi}{d\rho}\right)^{2}-\frac{2\cos^{3}\varphi}{\rho}\left(\frac{d^{2}\varphi}{d\rho^{2}}\right)\\ \notag
&&{}+\left[\frac{c_{0}^{2}}{2}-\frac{2c_{0}\sin\varphi}{\rho}+\frac{\lambda}{k}-\frac{\sin^{2}\varphi-2\cos^{2}\varphi}{2\rho^{2}}\right]\cos\varphi\left(\frac{d\varphi}{d\rho}\right)\\
&&{}+\left[\frac{\varDelta p}{k}+\frac{\lambda\sin\varphi}{k\rho}+\frac{c_{0}^{2}\sin\varphi}{2\rho}-\frac{\sin^{3}\varphi+2\sin\varphi\cos^{2}\varphi}{2\rho^{3}}\right].\notag
\\&&
\end{eqnarray}

\begin{figure}[hbpt]
  \centering
  \includegraphics[width=0.4\textwidth]{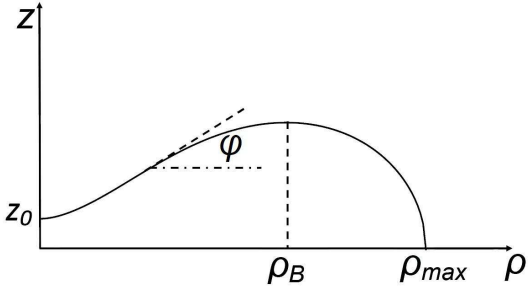}
  \caption{Cross section of RBC, only one quadrant is shown. There is rotational symmetry around the z axis and refiection symmetry at the $\rho$ axis.}
  \label{FIG1}
\end{figure}

In the same year, Naito \textit{et al.} found an analytical solution of (\ref{EQ3}) for $\Delta p$ = 0 and $\lambda$ = 0 as \cite{naito1993counterexample}
\begin{equation}\label{EQ4}
z=z_{0}+\int_{0}^{\rho}\tan\varphi\left(\rho_{1}\right)d\rho_{1},
\end{equation}
\begin{equation}\label{EQ5}
\sin\varphi=c_{0} \rho\ln\frac{\rho}{\rho_{B}},
\end{equation}
\noindent where as shown in FIG.\ref{FIG1}, $\rho$ is the distance of a point on the surface from the rotationally symmetric z-axis and $\varphi$ is the tangent angle of the point on the surface. $\rho_{B}$ is characteristic radius where $\varphi$ is zero, $\rho_{max}$ is the maximum value of $\rho$. When $\rho$ = $\rho_{max}$,
apparently $sin\varphi = -1$. $z_{0}$ is red blood cell's height when $\rho$ equals zero.
By fitting with the circular biconcave shape of RBC measured by Evans and Fung \cite{evans1972improved}, see FIG.3 in reference \cite{naito1996polygonal}, one finds the analytical solution (5) to exactly describe the RBC shape. Especially, the fitting reveals $c_{0}R_{0}=-1.67$ that confirms approximately the earlier prediction by Ou-Yang \textit{et al.} \cite{zhong1992determination}
\begin{equation}\label{EQ6}
c_{0}R_{0}=-1.62\approx-2/(\sqrt{5}-1).
\end{equation}

There is an immense size variation in vertebrate erythrocytes, even mammalian erythrocytes, which do not contain nuclei. Their size is different as they come from different species. So, we do dimensionless calculation to give a reasonable interpretation for human RBC and other mammalian erythrocytes in our paper. We do a series of transformation for the parameters which we mentioned above, $\rho^{'}_{max}=\rho_{max}/R_{0}$, $c^{'}_{0}=c_{0}R_{0}$, $\rho^{'}=\rho/R_{0}$ and $z_{0}^{'}=z_{0}/R_{0}$, where $R_{0}$ is the radius of sphere which has same area of RBC. Substitute these four parameters to (\ref{EQ5}) we get
\begin{equation}\label{EQ7}
-1=c_{0}^{'}\rho_{max}^{'}\ln\frac{\rho_{max}^{'}}{\rho_{B}^{'}},
\end{equation}
\noindent which yields a relational equation
\begin{equation}\label{EQ8}
\rho_{B}^{'}=\frac{\rho_{max}^{'}}{\exp\left(-\frac{1}{c_{0}^{'}\rho_{max}^{'}}\right)},
\end{equation}
\noindent substitute it back to (\ref{EQ5}) yields
\begin{equation}\label{EQ9}
\sin\varphi=c_{0}^{'}\rho^{'}\ln\frac{\rho^{'}}{\rho_{max}^{'}}-\frac{\rho^{'}}{\rho_{max}^{'}},
\end{equation}
\noindent when $\rho^{'}$ = $\rho^{'}_{max}$, we get the following equations from (\ref{EQ4})
\begin{equation}\label{EQ10}
z^{'}\left(\rho_{max}^{'}\right)=0=z_{0}^{'}+\int_{0}^{\rho_{max}^{'}}\tan\varphi\left(\rho^{'}\right)d\rho^{'},
\end{equation}
\noindent and
\begin{equation}\label{EQ11}
z_{0}^{'}=-\int_{0}^{\rho_{max}^{'}}\tan\varphi\left(\rho^{'}\right)d\rho^{'},
\end{equation}
\noindent Substitute it back to (\ref{EQ4}), we have
\begin{equation}\label{EQ12}
z^{'}\left(\rho^{'}\right)=-\int_{\rho^{'}}^{\rho_{max}^{'}}\tan\varphi\left(\rho^{''}\right)d\rho^{''}.
\end{equation}

Most mammalian erythrocytes are typically shaped as biconcave disks. First of all, this distinctive biconcave shape optimises the flow properties of blood in the large vessels, such as maximization of laminar flow and minimization of platelet scatter, which suppresses their atherogenic activity in those large vessels \cite{uzoigwe2006human}. Second, they are remarkably flexible and deformable so as to squeeze through tiny capillaries, where they efficiently release their oxygen load \cite{goodman2007human}. Third, take human red blood cells as a example, they take on average 20 seconds to complete one cycle of circulation, after that, most of them will have a ``body test'' in the spleen to guarantee themselves have enough deformability \cite{pivkin2016biomechanics}. Consideration of the number of all RBCs as quarter of total human cells, approximately estimate, 100 billion RBCs need to be produced and recirculated everyday. All these limited conditions have the same tendency that RBCs need to have a relatively smaller superficial area per unit volume. We calculate the surface area based on a basic principle that the volume of the RBC is constant and the biconcave disk shape is described by (\ref{EQ9}). Define the surface of RBC as a two-dimensional locus of points represented by a positional vector
\begin{equation}\label{EQ13}
\mathbf{r\left(\rho^{'},\theta\right)=\left(\rho^{'}\cos\theta,\rho^{'}\sin\theta,-\int_{\rho^{'}}^{\rho_{max}^{'}}\tan\varphi\left(\rho^{''}\right)d\rho^{''}\right)},
\end{equation}
\noindent where $\theta$ is azimuth angle. We have $\mathbf{r_{1}}$ and $\mathbf{r_{2}}$ respectively
\begin{eqnarray}\label{EQ14}
&&{}\notag \mathbf{r_{1}}=\frac{\partial\mathbf{r}}{\partial\rho^{'}}\\
&&=\left(\cos\theta,\sin\theta,-\frac{c_{0}^{'}\rho^{'}\ln\frac{\rho^{'}}{\rho_{max}^{'}}-\frac{\rho^{'}}{\rho_{max}^{'}}}{\sqrt{1-\left(c_{0}^{'}\rho^{'}\ln\frac{\rho^{'}}{\rho_{max}^{'}}-\frac{\rho^{'}}{\rho_{max}^{'}}\right)^{2}}}\right),
\end{eqnarray}
\begin{equation}\label{EQ15}
\mathbf{r_{2}}=\frac{\partial\mathbf{r}}{\partial\theta}=\left(-\rho^{'}\sin\theta,\rho^{'}\cos\theta,0\right).
\end{equation}
\noindent Then, the first fundamental form of the surface is obtained as
\begin{equation}\label{EQ16}
g_{\rho^{'}\rho^{'}}=\mathbf{r_{1}^{2}}=\frac{1}{1-\left(c_{0}^{'}\rho^{'}\ln\frac{\rho^{'}}{\rho_{max}^{'}}-\frac{\rho^{'}}{\rho_{max}^{'}}\right)^{2}},
\end{equation}
\begin{equation}\label{EQ17}
g_{\theta\theta}=\mathbf{r_{2}^{2}}=\rho^{'2},
\end{equation}
\begin{equation}\label{EQ18}
g_{\rho^{'}\theta}=\mathbf{r_{1}}\cdot\mathbf{r_{2}}=0,
\end{equation}
\begin{equation}\label{EQ19}
g\equiv\begin{vmatrix}
g_{\rho^{'}\rho^{'}} & g_{\rho^{'}\theta}\\
g_{\rho^{'}\theta} & g_{\theta\theta}
\end{vmatrix}=\frac{\rho^{'2}}{1-\left(c_{0}^{'}\rho^{'}\ln\frac{\rho^{'}}{\rho_{max}^{'}}-\frac{\rho^{'}}{\rho_{max}^{'}}\right)^{2}}.
\end{equation}
\noindent And the unit normal on the surface is derived as
\begin{eqnarray}\label{EQ20}
\mathbf{n}&=&\frac{\mathbf{r_{1}}\times\mathbf{r_{2}}}{\sqrt{g}}\notag \\
&=&\frac{1}{\sqrt{g}}(-\frac{c_{0}^{'}\rho^{'}\ln\frac{\rho^{'}}{\rho_{max}^{'}}-\frac{\rho^{'}}{\rho_{max}^{'}}}{\sqrt{1-\left(c_{0}^{'}\rho^{'}\ln\frac{\rho^{'}}{\rho_{max}^{'}}-\frac{\rho^{'}}{\rho_{max}^{'}}\right)^{2}}}\rho^{'}\cos\theta,\notag\\ &&\frac{c_{0}^{'}\rho^{'}\ln\frac{\rho^{'}}{\rho_{max}^{'}}-\frac{\rho^{'}}{\rho_{max}^{'}}}{\sqrt{1-\left(c_{0}^{'}\rho^{'}\ln\frac{\rho^{'}}{\rho_{max}^{'}}-\frac{\rho^{'}}{\rho_{max}^{'}}\right)^{2}}}\rho^{'}\sin\theta,\rho^{'}).
\end{eqnarray}
\noindent Therefore, we can calculate the equivalent surface area and equivalent volume of RBC respectively
\begin{eqnarray}\label{EQ21}
  S&=&2\int_{0}^{2\pi}\int_{0}^{\rho_{max}^{'}}\sqrt{g}d\rho^{'}d\theta \notag\\
  &=&4\pi\int_{0}^{\rho_{max}^{'}}\frac{\rho^{'}d\rho^{'}}{\sqrt{1-\left(c_{0}^{'}\rho^{'}\ln\frac{\rho^{'}}{\rho_{max}^{'}}-\frac{\rho^{'}}{\rho_{max}^{'}}\right)^{2}}},
\end{eqnarray}
\begin{eqnarray}\label{EQ22}
  V&=&\frac{2}{3}\int_{0}^{2\pi}\int_{0}^{\rho_{max}^{'}}\mathbf{r}\cdot\mathbf{n}\sqrt{g}d\rho^{'}d\theta \notag\\
    &=&4\pi\int_{0}^{\rho_{max}^{'}}\rho^{'}\notag\\
    &&\left(-\int_{\rho^{'}}^{\rho_{max}^{'}}\frac{c_{0}^{'}\rho^{''}\ln\frac{\rho^{''}}{\rho_{max}^{'}}-\frac{\rho^{''}}{\rho_{max}^{'}}}{\sqrt{1-\left(c_{0}^{'}\rho^{''}\ln\frac{\rho^{''}}{\rho_{max}^{'}}-\frac{\rho^{''}}{\rho_{max}^{'}}\right)^{2}}}d\rho^{''}\right)d\rho^{'}.\notag \\&&
\end{eqnarray}

\begin{figure}[hbpt]
  \centering
  \includegraphics[width=0.4\textwidth]{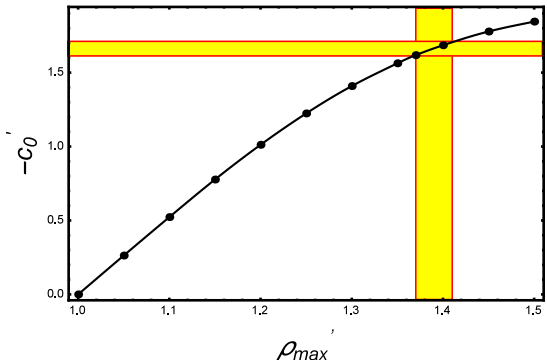}
  \caption{Minus equivalent spontaneous curvature $-c_{0}^{'}$ as a function of $\rho_{max}^{'}$}
  \label{FIGcp}
\end{figure}

Contour equation of RBC is decided only by two parameters $c_{0}^{'}$ and $\rho_{max}^{'}$ as described by (\ref{EQ9}) and (\ref{EQ11}). Under the condition that treats volume to be constant, we can numerically plot the curve of $-c_{0}^{'}$ as a function of $\rho_{max}^{'}$ with (\ref{EQ22}) as shown in FIG.\ref{FIGcp}.

\begin{figure}[hbpt]
  \centering
  \includegraphics[width=0.4\textwidth]{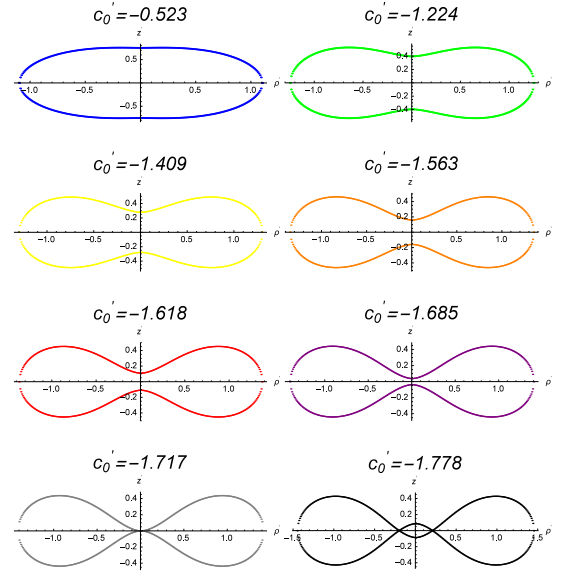}
  \caption{Series of RBC shape with different $c_{0}^{'}$}
  \label{FIG8countour}
\end{figure}

Using several sets of data ($c_{0}^{'}$¡¯, $\rho_{max}^{'}$) = (-0.523,1.1), (-1.012,1.2), (-1.224,1.25), (-1.409,1.3), (-1.618,1.37), (-1.685,1.4), (-1.717,1.414) and (-1.778,1.45), we draw series of RBC contour curves with (\ref{EQ9}) and (\ref{EQ12}) as shown in FIG.\ref{FIG8countour}.

  Obviously, The contour of RBC is most beautiful when $c_{0}^{'}$ equals -1.618, it seems the reason why the golden ratio should be chose, particularly we have a golden cross section as
\begin{equation}\label{EQ23}
\frac{\rho_{max}^{'}}{\rho_{B}^{'}}=\exp^{-\frac{1}{\rho_{max}^{'}c_{0}^{'}}}\thickapprox1.6.
\end{equation}

\begin{figure}[hbpt]
  \centering
  \includegraphics[width=0.4\textwidth]{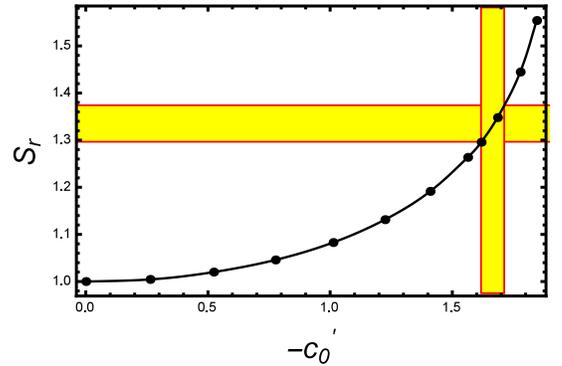}
  \caption{RBC's relative equivalent surface area as a function of -$c_{0}^{'}$}
  \label{FIGcsa}
\end{figure}

  However, we try to find more mechanism behind the beautiful shape. It is known spleen is the largest filter of RBCs in the body where the smallest openings for RBC passage are located. RBCs can be cleared from circulation when alterations in their size, shape, and deformability are detected. Surface area loss and reduced deformability will appear either physiological senescence or pathological alterations \cite{pivkin2016biomechanics}. From (\ref{EQ21}) and FIG.\ref{FIGcp} we plot the curve of relative equivalent surface area $S_{r}=S/4\pi$ as the function of equivalent spontaneous curvature -$c_{0}^{'}$ for illustration as shown in FIG.\ref{FIGcsa}.

\begin{figure}[hbpt]
  \centering
  \includegraphics[width=0.4\textwidth]{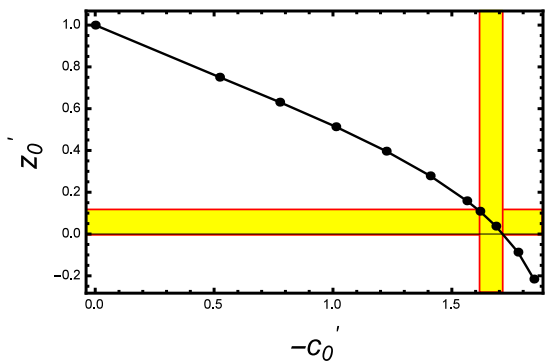}
  \caption{$z_{0}^{'}$ as a function of $-c_{0}^{'}$}
  \label{FIGcz}
\end{figure}

Apparently, the relative equivalent surface area $S_{r}$ increases monotonously with equivalent spontaneous curvature $-c_{0}^{'}$, thus, $-c_{0}^{'}$ decreases when the surface area gets loss. With (\ref{EQ11}) and FIG.\ref{FIGcp} one can plot the relation of $z_{0}^{'}$ and $-c_{0}^{'}$ as shown in FIG.\ref{FIGcz}

RBCs can pass the ``body test'' in spleen if the value of 2$z_{0}$ is no more than the length of the spleen's slit the average value which equals 0.65$\mu$m \cite{pivkin2016biomechanics}. According to $R_{0}$ = 3.25 $\mu$m \cite{zhong1992determination}, $z_{0}^{'}$ = $z_{0}/R_{0}$, and $z_{0}\geq 0$, we have $0 \leq z_{0}^{'} \leq 0.11$, then we have $1.618 \leq -c_{0}^{'} \leq 1.717$. All these allowed regions for $-c_{0}^{'}$, $\rho_{max}^{'}$ and $z_{0}^{'}$ have been colored by gold as shown in FIG.\ref{FIG8countour}, FIG.\ref{FIGcsa} and FIG.\ref{FIGcz}. Considering the allowed deviation of measurement, we can judge the golden ratio of $-c_{0}^{'}$ is the result of threshold of $z_{0}^{'}$ for RBCs to pass ``body test''. Under the perspective of saving cost in packing \cite{suzuki1996deformation,tsukada2001direct,skalak1989poiseuille,secomb1986flow}, the best choice is a sphere with $S_{r}$ = 1 , but RBCs cannot pass spleen, \textit{i.e.}, the deformability of the sphere is not enough. However, packing material will be wasted if $-c_{0}^{'} \geq 1.618$. Therefore, the golden ratio of $-c_{0}^{'}$ is the balance between the deformability and economical surface packing. In summary, we have found the mechanism behind the beautiful shape of RBCs.

\begin{acknowledgments}
We thank the National Basic Research Program of China (973 program, No. 2013CB932803) and the Joint NSFC-ISF Research Program, jointly funded by the National Natural Science Foundation of China and the Israel Science Foundation No. [51561145002] for financial support.
\end{acknowledgments}

\bibliographystyle{apsrev4-1}
\bibliography{zxj_thesis}

\end{document}